\documentclass[%
 aip,
 amsmath,amssymb,
 reprint,%
]{revtex4-1}

\usepackage{graphicx}
\usepackage{dcolumn}
\usepackage{bm}

\usepackage[utf8]{inputenc}
\usepackage[T1]{fontenc}
\usepackage{mathptmx}
\usepackage{etoolbox}
\usepackage{subfig}
\usepackage{pythonhighlight}

\makeatletter
\def\@email#1#2{%
 \endgroup
 \patchcmd{\titleblock@produce}
  {\frontmatter@RRAPformat}
  {\frontmatter@RRAPformat{\produce@RRAP{*#1\href{mailto:#2}{#2}}}\frontmatter@RRAPformat}
  {}{}
}%
\makeatother
\begin{document}

\preprint{AIP/123-QED}

\title{Self-supervised learning based on Transformer for flow reconstruction and prediction}
\author{Bonan Xu}
\affiliation{%
State Key Laboratory of Fluid Power and Mechatronic Systems, Department of Engineering Mechanics, Zhejiang University, Hangzhou 310027, P. R. China}
\author{Yuanye Zhou}
\affiliation{Baidu Inc., China}
\author{Xin Bian*}%
\affiliation{State Key Laboratory of Fluid Power and Mechatronic Systems, Department of Engineering Mechanics, Zhejiang University, Hangzhou 310027, P. R. China}
\email{bianx@zju.edu.cn}

\date{\today}

\begin{abstract}
Machine learning has great potential for efficient reconstruction and prediction of flow fields. However, existing datasets may have highly diversified labels for different flow scenarios, which are not applicable for training a model. To this end, we make a first attempt to apply the self-supervised learning (SSL) technique to fluid dynamics, which disregards data labels for pre-training the model. The SSL technique embraces a large amount of data ($8000$ snapshots) at Reynolds numbers of $Re=200$, $300$, $400$, $500$ without discriminating between them, which improves the generalization of the model. The Transformer model is pre-trained via a specially designed pretext task, where it reconstructs the complete flow fields after randomly masking $20\%$ data points in each snapshot. For the downstream task of flow reconstruction, the pre-trained model is fine-tuned separately with $256$ snapshots for each Reynolds number. The fine-tuned models accurately reconstruct the complete flow fields based on less than $5\%$ random data points within a limited window even for $Re=250$ and $600$, whose data were not seen in the pre-trained phase. For the other downstream task of flow prediction, the pre-training model is fine-tuned separately with $128$ consecutive snapshot pairs for each corresponding Reynolds number. The fine-tuned models then correctly predict the evolution of the flow fields over many periods of cycles. We compare all results generated by models trained via SSL and models trained via supervised learning, where the former has unequivocally superior performance. We expect that the methodology presented here will have wider applications in fluid mechanics.
\end{abstract}

\maketitle

\section{Introduction}
Machine learning-based models have made rapid progress in fluid mechanics~\cite{Brunton2020, Karniadakis2021, Zuo2023, Liang2022,Zhou2023}, mainly due to their rapidity, accuracy, and generalization for flow reconstruction and prediction. Compared to computational fluid dynamics~(CFD) methods, these models are more efficient in producing accurate results over a wide range of conditions~\cite{surrogate_model1, accelerated}. The superior performance is attributed to the incorporation of novel neural network architectures such as Transformer~\cite{Vaswani2017},
DeepONet~\cite{Lu2021a}, and Fourier Neural Operator~\cite{Li2021} among others, as well as a healthy software ecosystem and advances in GPU hardware. 

Typically, these models are trained using either data-driven~\cite{data_driven1, data_driven2, data_driven3, data_driven4} approaches or physical constraints~\cite{PINN, PINN1, PINN2, PINN3, PINN4}. When trained with abundant data via supervised learning, their accuracy is strongly dependent on the availability of labeled data of high quality~\cite{challenges1, challenges2, multi}. 
Thanks to advances in high-performance scientific computing,
extensive high-fidelity simulation data are indeed available for multiple flow scenarios~\cite{data1, data2, data3}.
Unfortunately, these datasets are not always directly applicable to supervised learning tasks due to their overwhelming number of labels, such as Reynolds numbers, Mach numbers, flow geometries, and so on, for a  wide range of flow scenarios.
In addition, merging datasets from different sources for transient flows is challenging due to variations in the label of the timestamp between snapshots in the datasets.
Thus, the data-driven branch of machine learning in fluid mechanics urgently needs innovative approaches to fully exploit the large number of multi-source datasets available, and possibly disregarding their labels for optimal utilization.

Contrary to data-driven methods, scientists have incorporated physical constraints, represented in partial differential equations~(PDEs), to guide the model training procedure in cases of insufficient or non-existent data.In particular, the physics-informed neural networks~(PINNs)~\cite{PINN, Raissi2020, Lu2021} have shown remarkable success. However, the incorporation of automatic differentiation for the PDEs leads to significant computational and memory overhead.
In addition, PINNs do not benefit from the aboundant data of diversified labels. 

This study aims to investigate the potential of self-supervised learning (SSL)~\cite{SSL1, SSL2} technique to address these critical challenges in fluid mechanics. Unlike traditional supervised learning, SSL leverages a vast amount of data and disregards their labels for pre-training,
which may not be directly related to the specific task being tackled. Subsequently, the pre-trained model is fine-tuned for various downstream tasks using only a limited amount of labeled data. Recently, influential frameworks such as Bert~\cite{bert}, Beit~\cite{beit}, and MAE~\cite{MAE} have utilized SSL to achieve significant success in natural language processing and computer vision. Inspired by these achievements, promising applications of SSL have already emerged in the fields of physics and biology~\cite{SSL_s1, SSL_s2, SSL_s3}. To the best of our knowledge, there has been no dedicated investigation of the technique in fluid mechanics. Consequently, this study represents the first attempt to apply SSL to flow reconstruction and prediction tasks.

With the progress in deep learning, there have been an increasing  efforts in flow reconstruction~\cite{reconstruction1, reconstruction2, reconstruction3} and prediction~\cite{prediction1, prediction2, prediction3}. Many studies have considered snapshots of the flow field as images, allowing them to leverage powerful techniques from computer vision, such as convolutional neural networks~(CNNs) and Vision Transformer. For instance, Laima et al. introduced DeepTRNet~\cite{DeepTRNet}, which is designed for time-resolved reconstruction of velocity fields around a circular cylinder. This approach incorporates a convolutional autoencoder to extract compact spatial representations embedded in the velocity field. Similarly, Xu et al. presented a super-resolution Transformer for turbulence~\cite{Super_resolution}, which enables reconstruction of turbulent flow fields with high quality. Additionally, Gao et al. employed physics-informed convolutional neural networks to achieve super-resolution and denoising of fluid flow~\cite{Gao}. However, these methods typically rely on data generated on a uniform grid or data generated on an unstructured grid but interpolated onto a uniform grid. To overcome the difficulties of processing data acquired directly from unstructured grids in CFD, researchers often resort to graph neural networks~(GNNs)~\cite{GNN1, GNN2, GNN3}. For example, Pfaff et al. presented the MeshGraphNets framework~\cite{MeshGraphNets}, which aims to master mesh-based simulations by using GNNs.

As a proof of concept, we generate a dataset for flows around a cylinder at moderate Reynolds numbers using CFD on unstructured grids. We employ the architecture of the 
Transformer~\cite{Vaswani2017} and further develop the Operator 
Transformer~\cite{li2022transformer}, which has been shown to work well with undecorated data  from unstructured grids.Within the Transformer, we implement a Galerkin-type attention mechanism~\cite{Galerkin} to guarantee discrete invariance, a crucial property in this particular context. To encode the absolute and relative positions of data points, we use the strategy of rotary position embedding~(RoPE)~\cite{Rope}. The main contribution of this study is the application of the Transformer-based SSL technique for flow field reconstruction and prediction. 
Specifically, during the pre-training phase, the Transformer model is trained with an extensive collection of unlabeled data covering a range of Reynolds numbers. To enhance the pre-training process,  we design a pretext task where the Transformer is trained to predict the complete flow fields after randomly masking $20\%$ data points in each snapshot.  In addition, we introduce a novel data augmentation method that maximizes the use of the available data. The pre-trained model then undergoes separate fine-tuning stages aimed at two downstream tasks, namely flow reconstruction and prediction.  To evaluate the SSL approach, we compare its performance with the same Transformer model trained by supervised learning. Our results demonstrate unequivocally that the SSL approach outperforms its counterpart for the two specified downstream tasks.

The remaining sections of the paper are organized as follows. Section~\ref{sec_method} presents the specific details of the SLL technique as well as a variant of the Transfomer architecture. In Section~\ref{sec_results}, intermediate results during the pre-trained stage and results of flow reconstruction and prediction as downstream tasks are presented. The results of the Transformer with and without the SSL technique are compared.  Section~\ref{sec_conclusion} provides a summary of the findings and suggests further studies of interest.

\section{Method and Dataset}
\label{sec_method}
In this section, we will introduce the procedure of SSL, the architecture of the Transformer neural networks, the data augmentation method, and the generation of the dataset.

\subsection{Self-supervised Learning}
\begin{figure}
\includegraphics[width=0.9\columnwidth]{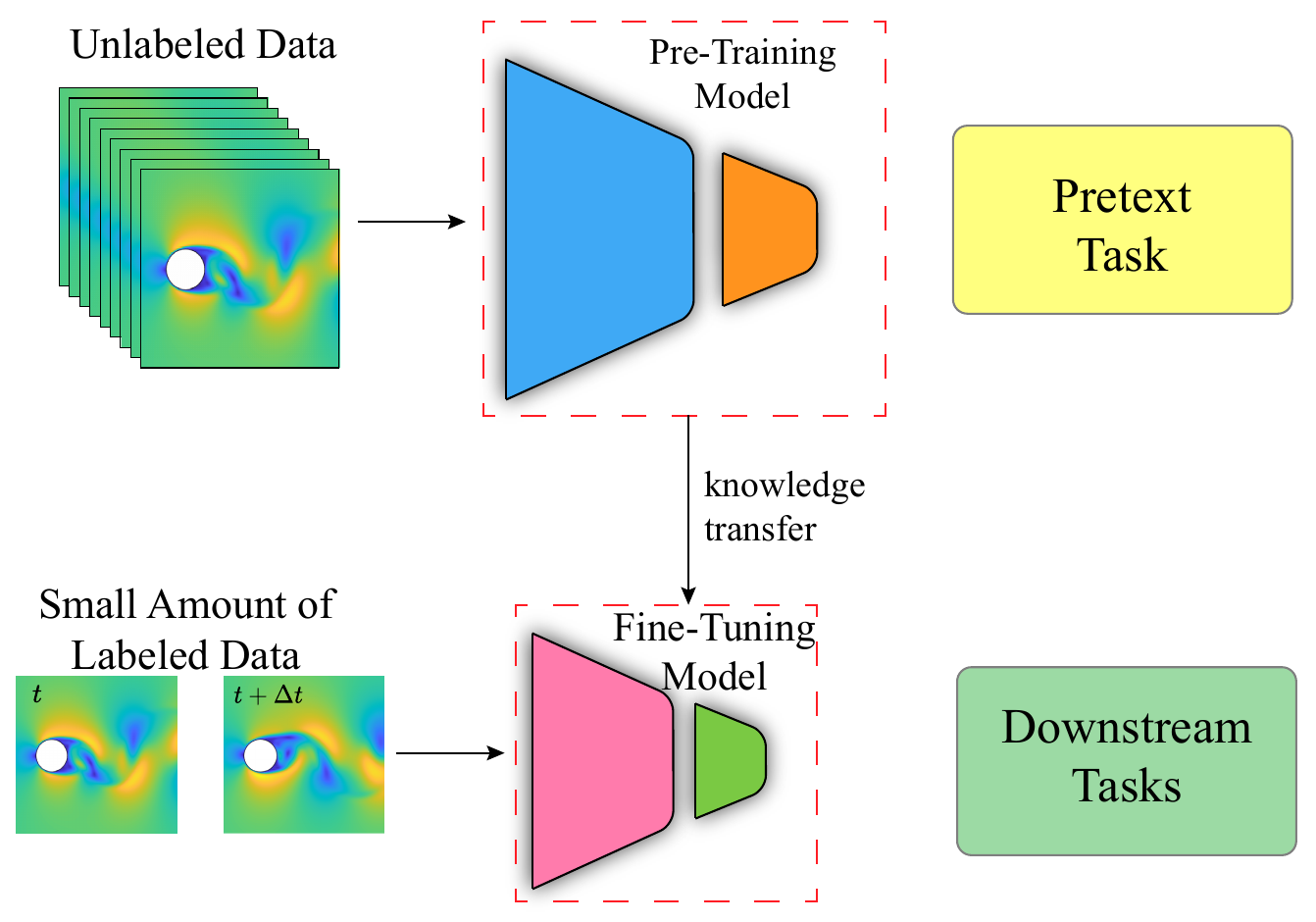}
\caption{\label{fig:sketch_ssl} Two stages of Self-Supervised Learning~(SSL) strategy. First, the model is trained with a pretext task using a large amount of unlabeled data during the pre-training phase, to enhance its generalization. Then, the model is fine-tuned individually for separate downstream tasks using a small amount of labeled data, to achieve a transfer learning.}
\end{figure}

SSL is an emerging machine learning technique that has been proposed to address the challenges of over-reliance on labeled data. Two primary stages of SSL in this study are illustrated in Fig.~\ref{fig:sketch_ssl}. First, during the pre-training phase, the model~(Transformer neural networks) is trained using a specifically designed pretext task. Then, an individually tailored fine-tuning is performed with a small amount of labeled data for each of the two different downstream tasks.

\subsubsection{Pretext Task}
The task designated for pre-training is referred to as pretext task. It is an essential part of  pre-training the model with unlabeled data, where the model generates its own pseudo-labels or supervisory signals based on the data themselves. The objective of pretext task is to guide to acquire a profound understanding of the inherent structures or patterns within the data governed by the Navier-Stokes equations,  which is crucial for the subsequent downstream tasks of practical interest.

Specifically, the input to encoder of the pre-trained model consists of incomplete snapshots of the flow field along with their corresponding coordinates. Similar to BERT~\cite{bert}, a benchmark work in SSL for natural language processing, approximately 20\% of the flow data points are randomly masked in each snapshot. The neural networks are then trained to reconstruct the entire flow field, including pressure and velocity, based on the coordinates provided to the decoder.
 
The pretext task thus designed empowers the model to leverage diverse data during the pre-training phase, where all $N_{tot}$ snapshots at different Reynolds numbers can be employed for the training. By incorporating a large amount of unlabeled data, this strategy significantly improves the model generalization.

\subsubsection{Downstream Tasks}

We identify two different types of downstream tasks to illustrate the effectiveness of the SSL strategy in improving model performance.  

The first type of downstream tasks involves fine-tuning the pre-trained model using a small amount of labeled data, i.e., $N_1$ snapshots relevant to the task. The model is then used to predict the full flow field based on the sparse data available within a limited window around the cylinder.  It is worth noting that the given data points are not constrained to specific positions, and their indices within the input array are subject to random sampling. In practical applications of experiments, this capability can be valuable when one can only acquire limited and randomly distributed data within a small observation window.

For the second type of downstream tasks, the pre-trained model is fine-tuned by a small amount of labeled data, i.e., $N_2$ snapshot pairs relevant to the task.  The snapshot pairs are randomly sampled from the dataset, with a constant time interval between the two snapshots in the pair. This fine-tuning process aims to create a surrogate model capable of predicting the flow field in the future, i.e., based on any input snapshot, it generates an output snapshot at a later time. This process is repeated to predict the flow fields over a long period of time.

\subsection{Transformer neural networks}

As discussed in the above section, the SSL strategy introduces additional requirements on the architecture of the neural networks. Firstly, it is crucial to avoid introducing index bias to the input data during the training process.  Therefore, the index of each data point is randomly assigned with positional information obtained from a positional embedding. Secondly, it must be able to accommodate inputs of variable lengths, as some data points are randomly excluded from the input arrays during the pre-training phase. This attribute is commonly refered to as the discrete invariance of the model. Lastly, it should be able to provide values at any given query location, as a consequence of the prevalent utilization of unstructured grids by CFD for complex geometries.  To meet the last two requirements, we shall further modify the operator Transformer originally proposed by Li et al~\cite{li2022transformer}.

\subsubsection{Main Architecture}

\begin{figure}
\includegraphics[width=0.9\columnwidth]{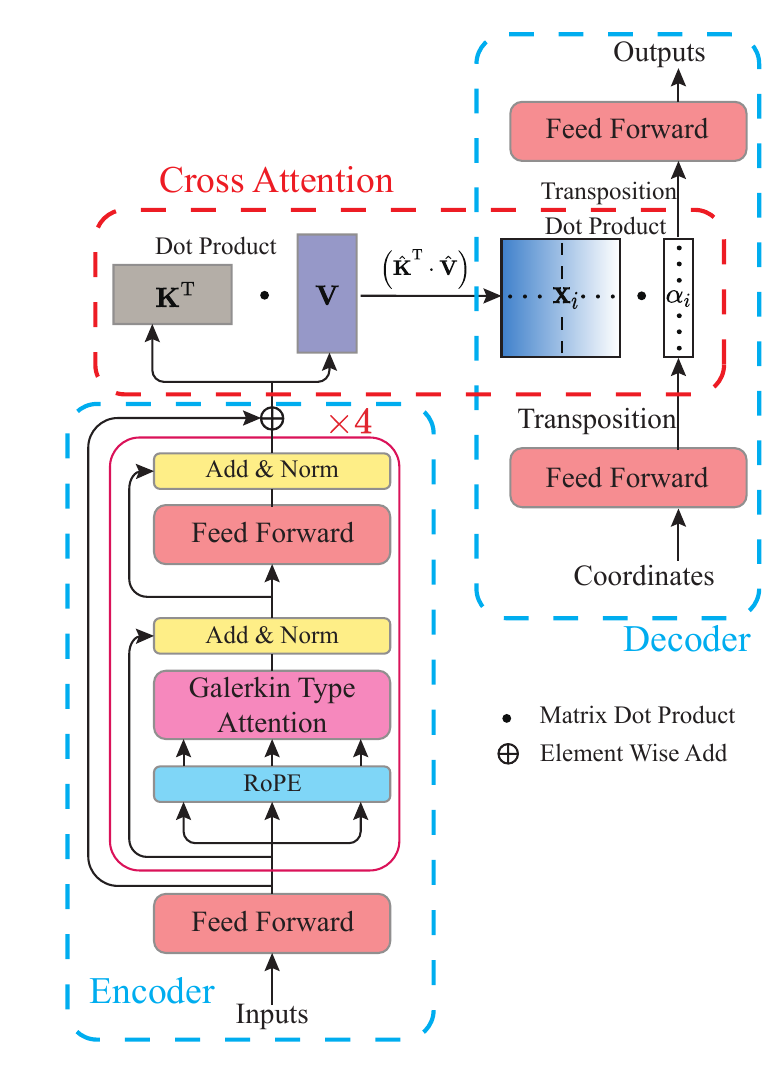}
\caption{\label{fig:sketch_transformer} The primary architecture of a variant of the Transformer.}
\end{figure}

The primary architecture of a variant of the Transformer is shown in Fig.~\ref{fig:sketch_transformer}, where an encoder and an decoder\cite{En_De1} are concatenated. The encoder consists of a feedforward neural network~(FNN) as input embedding and a stack of $N=4$ identical attention modules. The FNN is a simple stacking of two point-wise linear layers with GELU activation function to map the inputs $\mathbf{y}_{in}$ from a low dimensional physical space to the feature vectors $\mathbf{y}$ in feature space $\mathbb{R}^{d}$. The expansion ratio of the hidden channel is set to be $2$. This process can be expressed as
\begin{equation}
   \text{FNN}(\mathbf{y}) = \mathbf{W}_{2}\cdot(\text{GELU}(\mathbf{W}_{1}\cdot\mathbf{y}_{in} + \mathbf{b}_{1})) + \mathbf{b}_{2}.
\end{equation}
Each attention module is composed of one positional embedding layer and two sub-layers. The first sub-layer of the attention module employs a multi-head Galerkin-type self-attention mechanism, while the second sub-layer consists of a pointwise FNN. Residual connections are utilized around each sub-layer, and layer normalizations (LN) are applied both before and after the attention layer. 
 
Since the input data points are seen as a long sequence in the Transformer architecture, it becomes imperative to incorporate the positional information to the corresponding data points. Both absolute and relative positions hold significance within fluid systems. Following a similar approach as in the previous research~\cite{li2022transformer}, we use the rotary position embedding (RoPE)~\cite{Rope} strategy. This embedding technique was initially introduced by Sun et al.~\cite{Rope} and subsequently extended to multidimensional systems in a blog post authored by Biderman et al~\cite{rope-eleutherai}. For an in-depth understanding of the formulation, please refer to the details provided in the Appendix. The entire operation within the attention module can be articulated as follows:
\begin{align} 
\hat{\mathbf{y}}^{l} &= \text{RoPE}(\mathbf{y}^{l}),  \\
\mathbf{y}^{l+ \frac{1}{2}} &= \text{LN} \left( \text{Attention}(\text{LN}(\hat{\mathbf{y}}^{l})) + \mathbf{y}^{l} \right), \\
\mathbf{y}^{l+1} &= \text{FFN}(\mathbf{y}^{l+ \frac{1}{2}}) + \mathbf{y}^{l+ \frac{1}{2}},
\end{align}
where $\mathbf{y}^{l}$ represents the output from the preceding layer or module
and $\hat{\mathbf{y}}^{l}$ corresponds to the vector after the application of positional embedding. Furthermore, $\mathbf{y}^{l+1}$ denotes the output of the second sub-layer and serves as the input for the subsequent layer or module.

Within the decoder, the inputs consist of coordinates, denoted as $\mathbf{X}$, representing the proposed locations. To enable the Transformer to reconstruct/predict flow quantities at any given location, all operations in the decoder are conducted on a point-wise basis. This approach ensures that the output of decoder is independent of the positions of other points, relying solely on the coordinate information of the individual proposed points.  These coordinates $\mathbf{X}$ are projected into the feature vectors $\mathbf{F}$ within the feature spaces through the coordinate projection module. This module employs a concatenation of Fourier feature mapping~\cite{Fourier_features} and FNN. By passing input points through a simple Fourier feature mapping, the FNNs can learn the high-frequency functions in low-dimensional problem domains more effectively ~\cite{Fourier_features}.

The Galerkin-type cross-attention mechanism is employed to incorporate information from the input sequence into the layers of the decoder. Finally, the internal vectors are transformed into flow fields through a pointwise FNN, which serves as the outputs of the complete operator.

\subsubsection{Galerkin Type Attention Mechanism}

\begin{figure}[htbp]
	\centering
	\subfloat[Classical attention mechanism]{\includegraphics[width=.9\columnwidth]{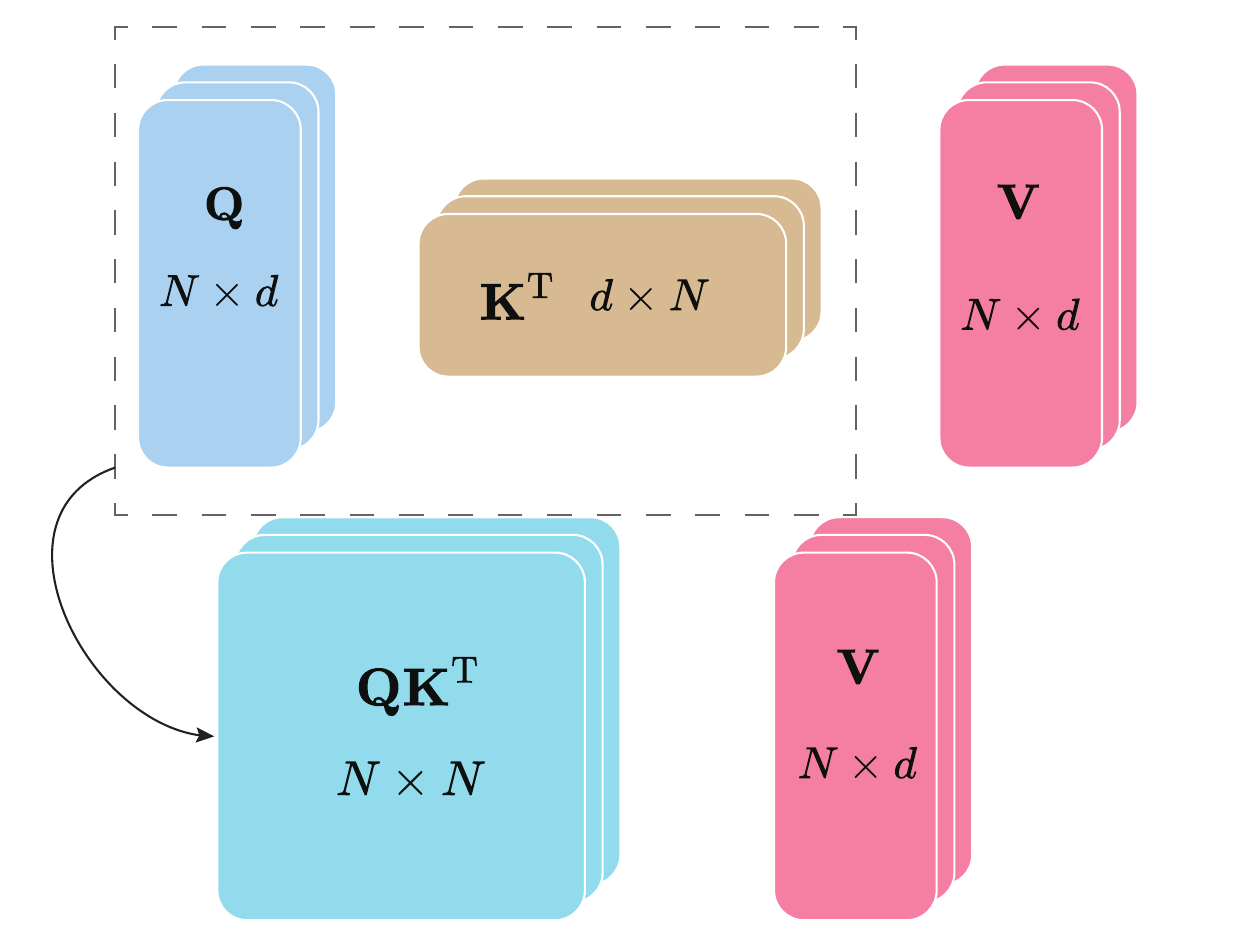}
    \label{fig:sketch_attention_classical}} \\
	\subfloat[Galerkin-type attention mechanism]{\includegraphics[width=.9\columnwidth]{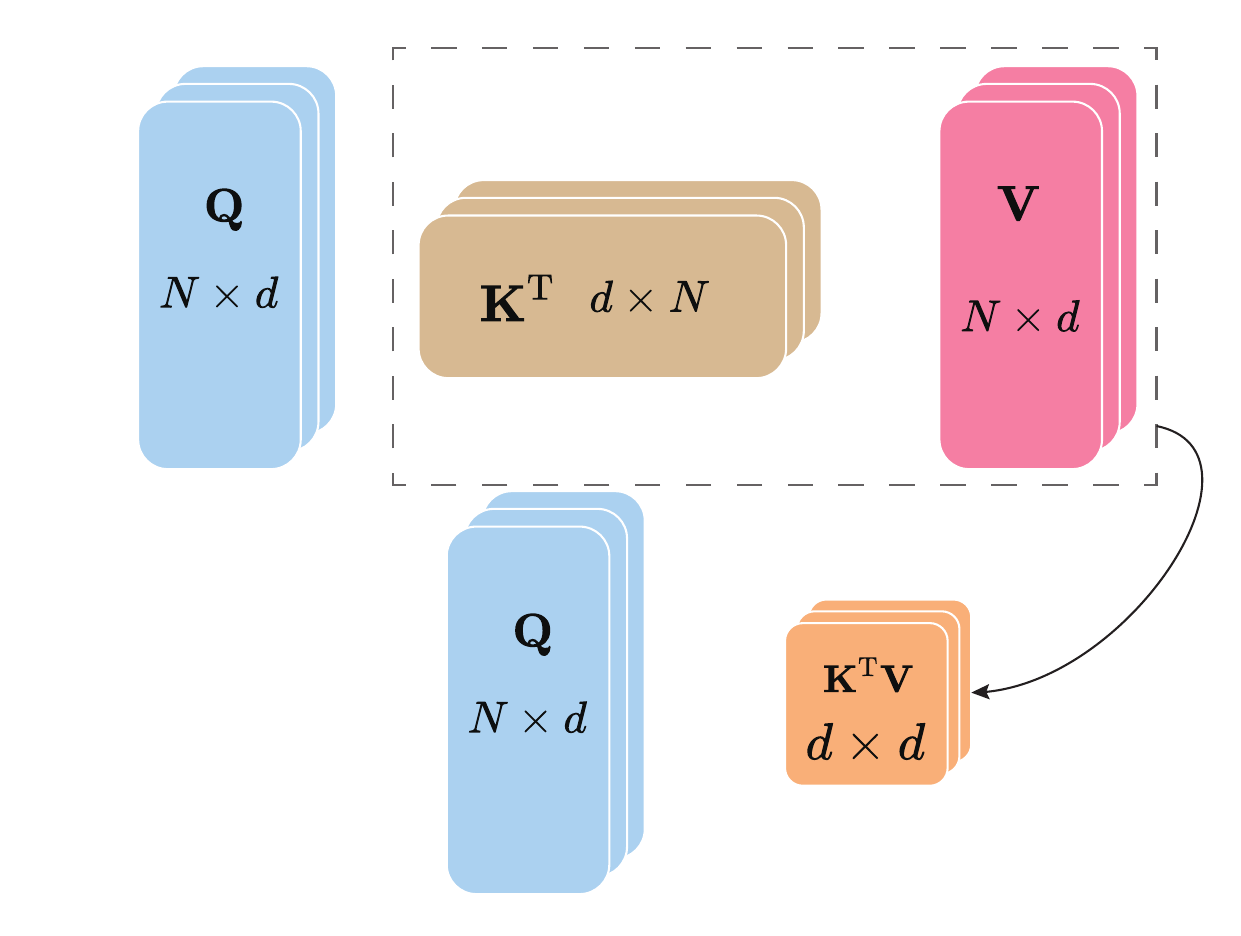}
    \label{fig:sketch_attention_galerkin}}
	\caption{The structures of the classical attention mechanism and the Galerkin-type attention mechanism.}
    \label{fig:sketch_attention}
\end{figure}

The detailed structures of the classical attention mechanism and the Galerkin-type attention mechanism are shown in Fig.~\ref{fig:sketch_attention}. In the classical attention, as shown in Fig.~\ref{fig:sketch_attention_classical}, the input $\mathbf{y} \in \mathbb{R}^{N \times d}$ from previous layer is mapped to query $\mathbf{Q}$, key $\mathbf{K}$, and value $\mathbf{V}$ matrices in $\mathbb{R}^{N\times d}$ space. The output of attention layer is computed by the dot product between attention map $\text{softmax}\left( \frac{\mathbf{QK}^{\mathrm{T}}}{\sqrt{ d_{k} }} \right)$ and value vector $\mathbf{V}$. The detailed process can be written as:
\begin{align}
&\mathbf{Q} =  \mathbf{y}\mathbf{W}^{Q}, \quad \mathbf{K} =  \mathbf{y}\mathbf{W}^{K}, \quad \mathbf{V} =  \mathbf{y}\mathbf{W}^{V}, \\
&\text{attention}(\mathbf{Q}, \mathbf{K}, \mathbf{V}) = \text{softmax} \left( \frac{\mathbf{QK}^{\mathrm{T}}}{\sqrt{ d_{k} }} \right) \mathbf{V},
\end{align}
where $\mathbf{W}^{Q}$, $\mathbf{W}^{K}$, and $\mathbf{W}^{V}$ are the weight matrices to be trained.
This form of attention mechanism has demonstrated significant successes across various fields. However, its implementation can also result in a substantial computational and memory overhead, especially for a system of complex fluid dynamics with a large number of solution points.

Instead, we employ the Galerkin-type attention mechanism developed by Cao\cite{Galerkin}. The structure of the Galerkin-type attention is depicted in Fig.~\ref{fig:sketch_attention_galerkin}.
It can be formulated as follows: 

\begin{align}
& \mathbf{Q} =  \mathbf{y}\mathbf{W}^{Q}, \quad \mathbf{K} =  \mathbf{y}\mathbf{W}^{K}, \quad \mathbf{V} =  \mathbf{y}\mathbf{W}^{V}, \\
& \text{attention}(\mathbf{Q}, \mathbf{K}, \mathbf{V}) = \frac{1}{n} \mathbf{Q}\left( \hat{\mathbf{K}}^{\mathrm{T}} \hat{\mathbf{V}} \right),
\end{align}
where $\hat{\cdot}$ represent a column-wise normalized (via instance normalization) matrix.
This kind of attention can achieve discrete invariance, because the number of data points, denoted as $N$, will be eliminated in the dot product between matrix $\hat{\mathbf{K}}^{\mathrm{T}} \in \mathbb{R}^{d\times N}$ and matrix $\hat{\mathbf{V}} \in \mathbb{R}^{N \times d}$ , yielding a resulted matrix with dimensions $[d \times d]$. 
If the self-attention mechanism was contained in the decoder,
the matrix $\mathbf{Q}$, $\mathbf{K}$ and $\mathbf{V}$ would be supplied by encoder or decoder itself.
However, in cross-attention layers, the query matrix $\mathbf{Q}$ is sourced from the decoder, while the encoder provides the Galerkin-type attention map denoted as $\left( \hat{\mathbf{K}}^{\mathrm{T}} \hat{\mathbf{V}} \right)$. Mathematically, the Galerkin-type attention map can be regarded as a collection of learnable basis vectors. The outputs of the Galerkin-type attention mechanism can be seen as a linear combination of these basis vectors, with the learnable parameters $\mathbf{Q}$ determining the weights of the combination.

\subsubsection{Data Augmentation}

We introduce a new data augmentation method tailored for Transformer networks. Unlike convolutional neural networks, Transformer networks derive position information from positional embedding. Consequently, the index of the data point within the input sequence can be randomly selected, as long as the corresponding coordinate information is embedded. Furthermore, given the discrete invariance property inherent to the Transformer, it becomes feasible to randomly remove certain data points, thereby inducing variations in the input sequence length.  Using these data augmentation strategies, the input sequences of the same snapshot may not be exactly the same in different epochs.  This data augmentation strategy ensures that the Transformer operator genuinely learns the relationship between data points, rather than solely focusing on the order of data points within the input sequence. This deliberate diversification contributes to an enhanced network generalization capability. This data augmentation can be implied in a few lines of code in Python:

\begin{python}
# x: input sequence
length = x.shape[-2]
index = random.sample(range(0, length-1),
                    length)
x = x[:, :, index, :]
\end{python}

\subsubsection{Dataset}

\begin{figure}[h]
\includegraphics[width=0.95\columnwidth]{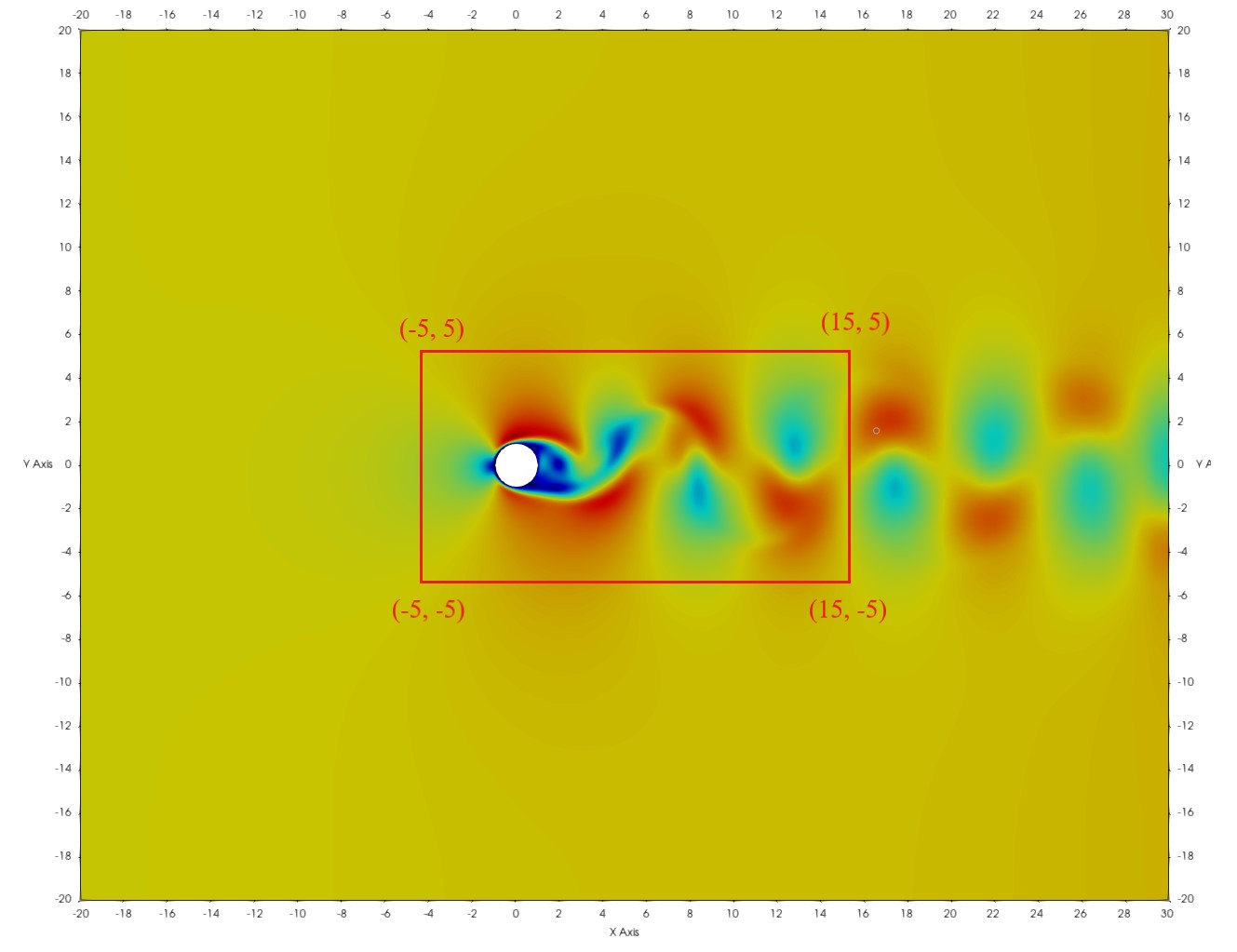}
\caption{\label{fig:data_geo} Simulation data within a windom of size $[-5, 15] \times [-5,5]$ around the cylinder are sampled every $50\delta t$ at steady states as snapshots to populate the dataset.}
\end{figure}
The dataset is generated from 2D time-dependent flows around a cylinder. The governing equations are the incompressible Navier-Stokes equations,
\begin{align}
& \nabla\cdot \mathbf{u} = 0, \\
& \frac{\partial \mathbf{u}}{\partial t} + (\mathbf{u} \cdot \nabla) \mathbf{u} = -\frac{1}{\rho} \nabla p + \nu \nabla^2 \mathbf{u},
\end{align}
where $\mathbf{u}$, $t$, $\rho$, $p$ and $\nu$ represent the velocity, time, density, pressure, and kinematic viscosity, respectively. On the left of the computational domain is the uniform inflow boundary with velocity magnitude $U$ and on the right is the outfow boundary with zero presure gradient. Peridoic boundary conditions are applied at the top and bottom. The Reynolds numbers $Re=U a/\nu$ are $200$, $250$, $300$, $400$, $500$, and $600$ with $a$ being the radius of the cylinder. All simulations are conducted with the open-source OpenFOAM library with $368,000$ points on unstructured grids, with denser distribution of points around the cylinder. The time step is set universally as $\delta t = 2\times 10^{-2}s$.

To focus on the variation of complex flows, we customize a window of size $[-5, 15] \times [-5,5]$ around the cylinder as illustrated in Fig.~\ref{fig:data_geo}, which contains $14,914$ grid points.
The dataset is constructed by sampling flow field data within this window at a constant time interval $\Delta t = 50\delta t = 1s$ of steady states. The corresponding configuration files for the simulations can be found in supplementary information, by which one should be able to generate the same dataset.

\begin{figure}[h]
\includegraphics[width=0.95\columnwidth]{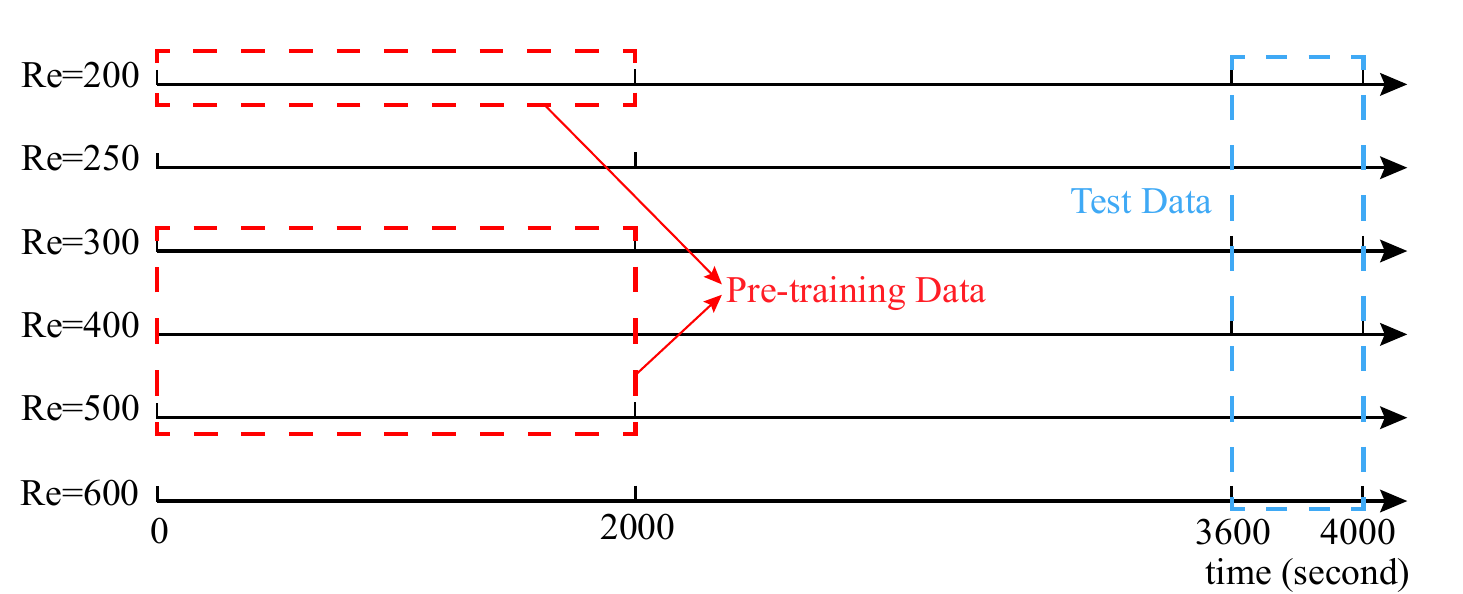}
\caption{\label{fig:data_split} Division of data into pre-training set and test set. At time $0$ of the dataset, the simulations are already at steady states. }
\end{figure}

The dataset splits into pre-training set and testing set, as shown in Fig.~\ref{fig:data_split}, where the first $2000$ snapshots of each Reynolds number are taken as the pre-training set, and all the last $400$ snapshots are reserved as the test set. Data for $Re=250$ and $600$ are deliberately excluded from the pre-training set to examine the interpolation and extrapolation performance of the model.Moreover, a small number of snapshots are randomly chosen from the first $2000$ ones and they will be employed via supervised learning as labled data, to fine-tune the pre-trained model for specific downstream tasks.

\section{Results and Discussion}
\label{sec_results}
In our numerical experiments, the model dimension and feature dimension after the input embedding are set to be $128$. There is only one attention head. The total number of trainable parameters in the Transformer is $1,268,355$, which is devided into $821,763$ for the encoder and $446,592$ for the decoder.

The core process of SSL comprises of a pretext task during the pre-training phase and fine-tunings towards specific downstream tasks. The step-by-step procedure is outlined as follows:
\begin{itemize}
    \item Randomly initialize the parameters of the Transformer network.
    \item Pre-train the model iteratively using $N_{tot}$ snapshots disregarding their labels (that is, Reynolds numbers and timestamps) so that both the training and test errors reduce to small magnitudes.
    \item Store the network's parameters as the primed state of the pre-trained model.
    \item Load the parameters of the pre-trained model and fine-tune it via supervised learning using $N_1$ snapshots from a specific Reynolds number.
    \item Given sparse data points at any instant, accomplish the task of flow reconstruction for the entire flow field.
    \item Load the parameters of the pre-trained model and fine-tune it via supervised learning using $N_2$ pairs of snapshots from a specific Reynolds number.
    \item Given one arbitrary starting snapshot, accomplish the task of flow prediction over next snapshots.

\end{itemize}

\subsection{Pretext Task}
During the pre-training stage, the snapshots collected at Reynolds number of $200$, $300$, $400$, and $500$ are utilized for training and there are $N_{tot}=2000\times 4=8000$ snapshots in total, as illustrated in Fig.~\ref{fig:data_split}. Accordingly, $400\times 4=1600$ snapshots, excluding data for $Re=250$ and $600$, in the test set are utilized for testing the efficacy of the pre-traning.

\begin{figure}[h]
\includegraphics[width=0.95\columnwidth]{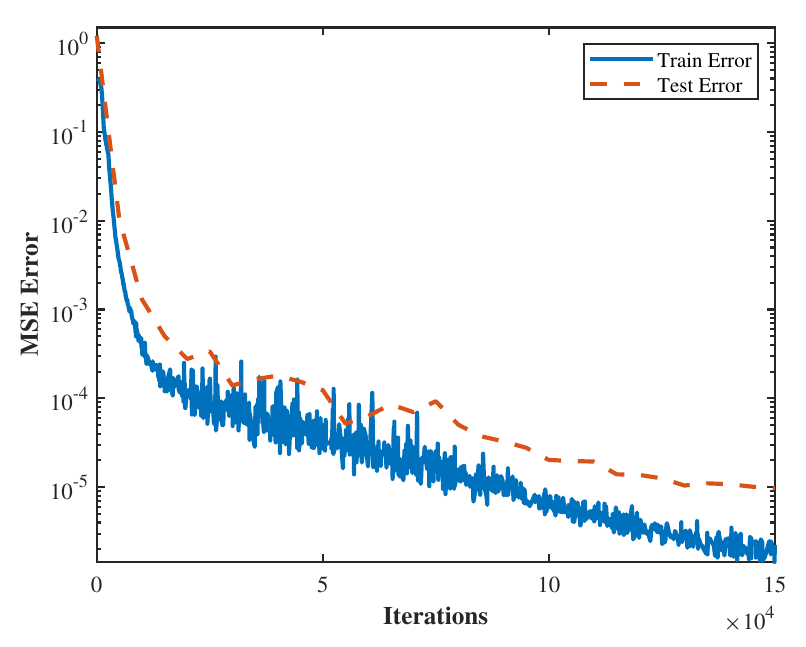}
\caption{\label{fig:mse_pretrain} Train error and test error expressed as mean square error~(MSE) in the pre-training stage.}
\end{figure}
For each snapshot in the pre-training set, $20\%$ data points are randomly masked and the Transformer is tasked to predict the complete snapshot based on the $80\%$ data points available. That is, $80\%$ data points of $(x_1, x_2, u_1, u_2, p)$ in each snapshot are absorbed to the encoder and $100\%$ coordinates ($x_1, x_2)$ are provided to the decoder so that a complete snapshot is predicted. The training error is computed as the difference between the predicted values and the true values over the complete snapshots in the pre-training set. The test error is computed similarly, but on the data in the test set. 

The parameters of the Transformer network are optimized towards minimizing the training error expressed as mean squared error~(MSE) via the AdamW algorithm~\cite{adamw}, with a weight decay coefficient of $10^{-4}$. The learning rate is managed following a one-cycle policy~\cite{one_cycle}: starting with an initial learning rate of $6\times 10^{-7}$, reaching a maximum of $6 \times 10^{-4}$, and then decreasing to a final rate of $6\times 10^{-8}$. The proportion of the cycle spent at increasing the learning rate is set as $0.3$. Data are processed in batches of size $16$ snapshots, and the optimization process terminates after $150,000$ iterations.

In Fig.~\ref{fig:mse_pretrain}, it is shown that both the training and test losses decrease as the number of training iterations increases. After the inital sharp decrease, both the training and test losses display pronounced oscillations. This behavior is mainly due to the utilization of the warm-up method to adapt the learning rate. In particular, the learning rate is maximal during this stage, which helps the model to get rid of the local minimal solutions. After a long period of iterative training, the training loss and test loss eventually diminish to the order of $10^{-6}$ and $10^{-5}$, respectively. Overall, the difference between the training error and testing error remains small. This result underscores the effectiveness of the proposed pretext task, which approves the utilization of an expanded dataset for training the network and thereby, mitigates overfitting while enhances the generalization capabilities of the model. Consequently, the parameters of the pre-trained model are stored, which are the foundation for subsequent fine-tunings of the model aimed at individual downstream tasks.

\subsection{Downstream Task 1: Reconstruct flow field from limited and randomly located data}

\begin{figure}[h]
\includegraphics[width=0.95\columnwidth]{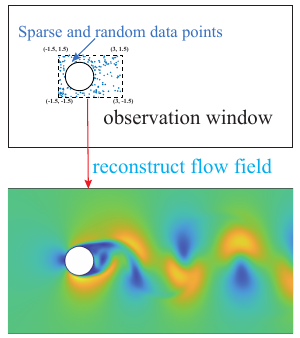}
\caption{\label{fig:sketch_re} Sketch for reconstruction of flow field. To reconstruct the the complete flow field, only a small portion of data points at random locations within the observation window are available, which amounts to $2.37\%$ to $4.75\%$ of the test data points. }
\end{figure}

In practical experiments and applications, it may turn up that only a sparse amount of data can be acquired, as illustrated in Fig.~\ref{fig:sketch_re}, where a small observation window around the cylinder contains sparse and random data points. Moreover, neither the quantity nor the distribution of the data remain consistent between snapshots. In such scenarios, the task of reconstructing the complete flow field of each snapshot becomes extremely challenging. Neural networks lacking discrete invariance are inclined to interpolate input data to a pre-determined length, which often introduces significant errors for sparse data.

To accomplish the task of flow reconstructions at six Reynolds numbers, fine-tunings of the pre-trained model are seperately carried out via supervised learning. For each $Re$, $N_1=256$ snapshots are randomly selected among the first $2000$ ones as labeled data. It is worth noting that for $Re=200$, $300$, $400$ and $500$, the snapshots were already used for the pre-training, while for $Re=250$ and $600$, the snapshots are used for the first time. Specifically, $n$ data points with five dimensions $(x_1, x_2, u_1, u_2, p)$ within the observation window, as illustrated in Fig.~\ref{fig:sketch_re}, are absorbed by the encoder. Here $n$ is a random number sampled from $[354, 708]$, which amounts to only $2.37\%$ to $4.75\%$ of all data points of one complete snapshot. Meanwhile, $100\%$ coordinates ($x_1, x_2)$ are provided to the decoder so that a complete snapshot is reconstructed by the Transformer model. The training error is computed as the difference between the predicted values and the true values over the entire $256$ snapshots.

Analogous to the pretext tasks, the learning rate in these fine-tunings follows to the one-cycle learning rate policy. The policy is governed by initial, maximum, and final learning rates set at $3\times10^{-7}$, $3\times10^{-4}$, and $3\times10^{-7}$, respectively. The total number of iterations spans $60,000$.  All other training configurations remain consistent with those employed in the pretext task.

\begin{figure*}
    \centering
    \subfloat[MSE error at Reynolds number $200$]{\includegraphics[width=.65\columnwidth]{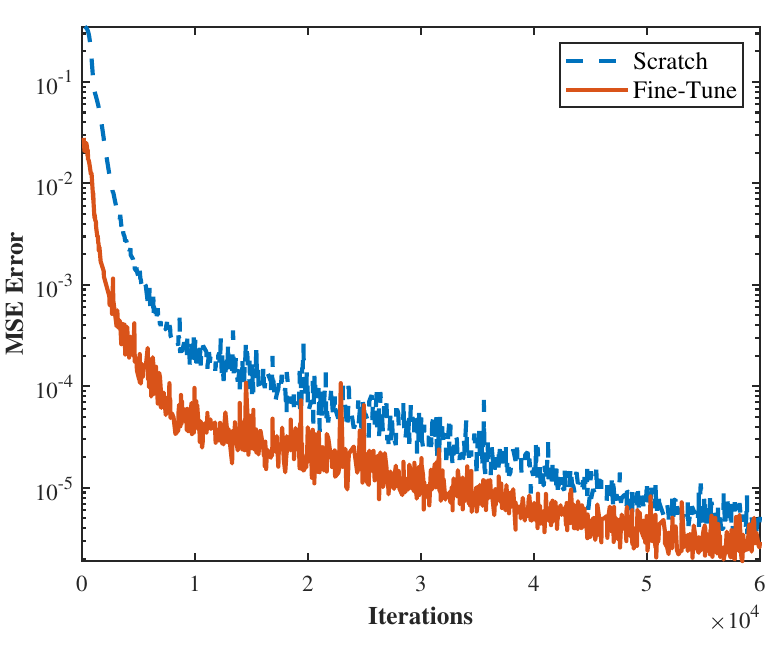}}\hspace{1pt}
    \subfloat[MSE error at Reynolds number $250$]{\includegraphics[width=.65\columnwidth]{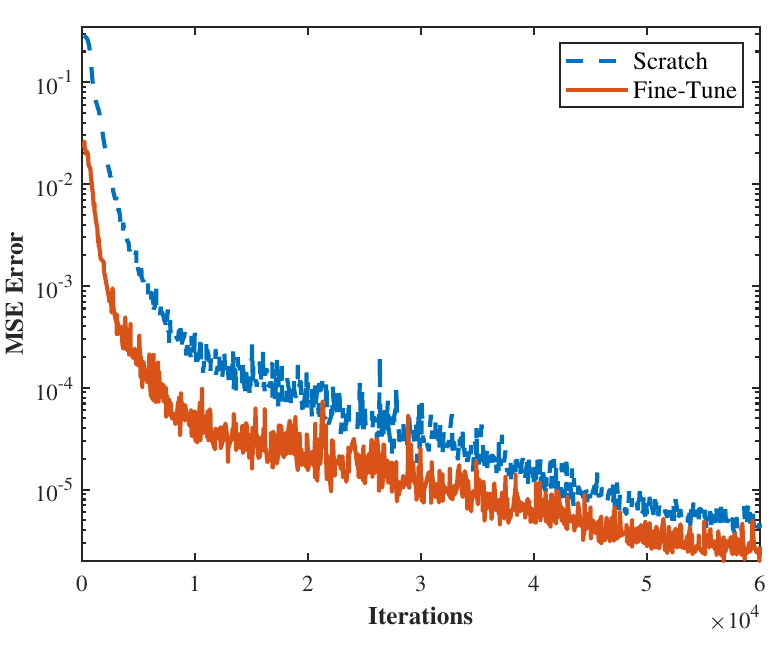}}\hspace{1pt}
    \subfloat[MSE error at Reynolds number $300$]{\includegraphics[width=.65\columnwidth]{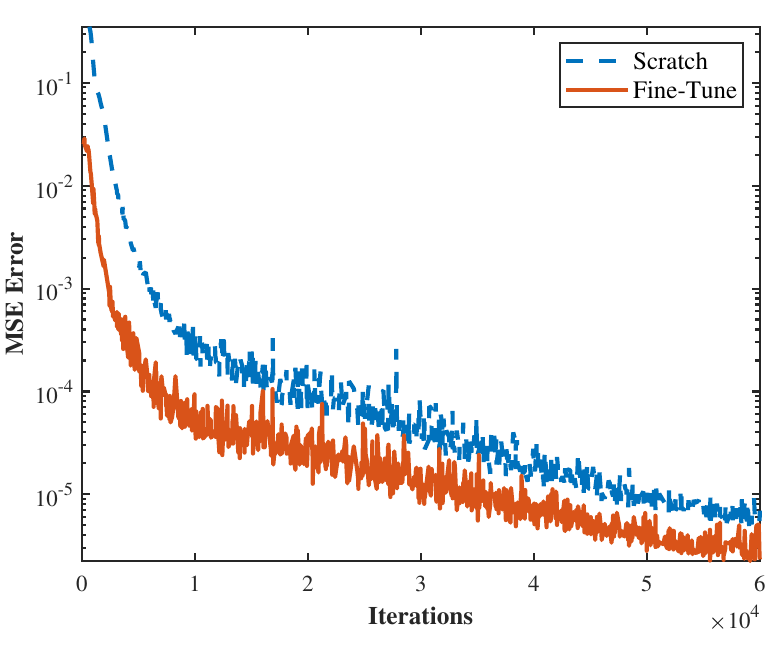}}\\
    \subfloat[MSE error at Reynolds number $400$]{\includegraphics[width=.65\columnwidth]{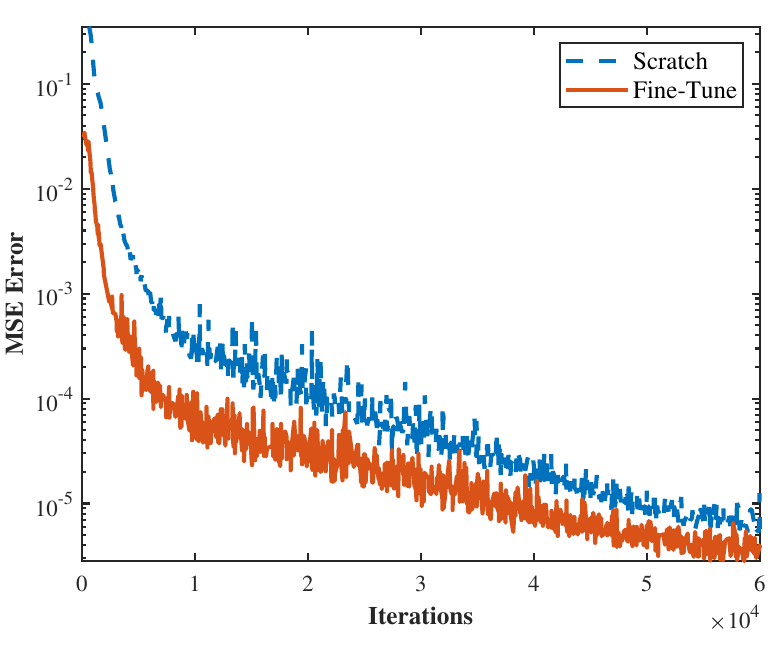}}\hspace{1pt}
    \subfloat[MSE error at Reynolds number $500$]{\includegraphics[width=.65\columnwidth]{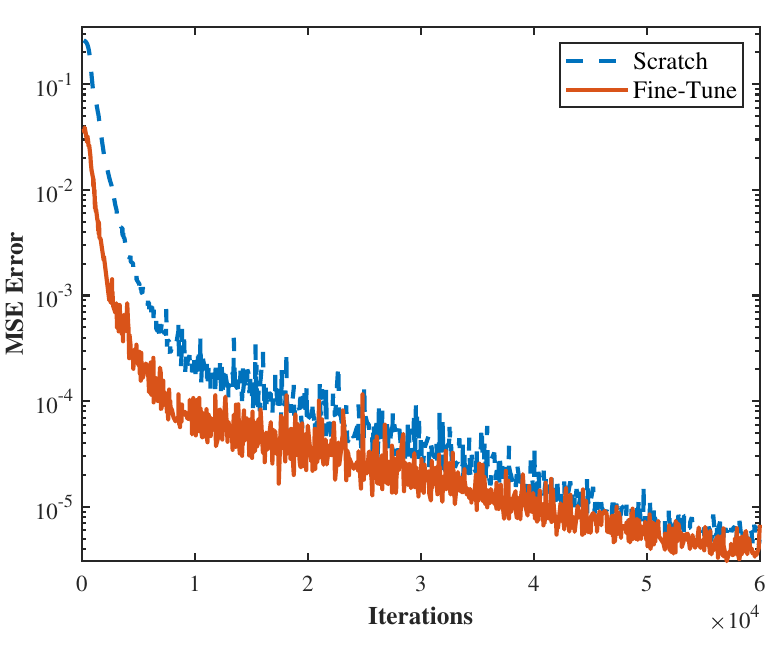}}\hspace{1pt}
    \subfloat[MSE error at Reynolds number $600$]{\includegraphics[width=.65\columnwidth]{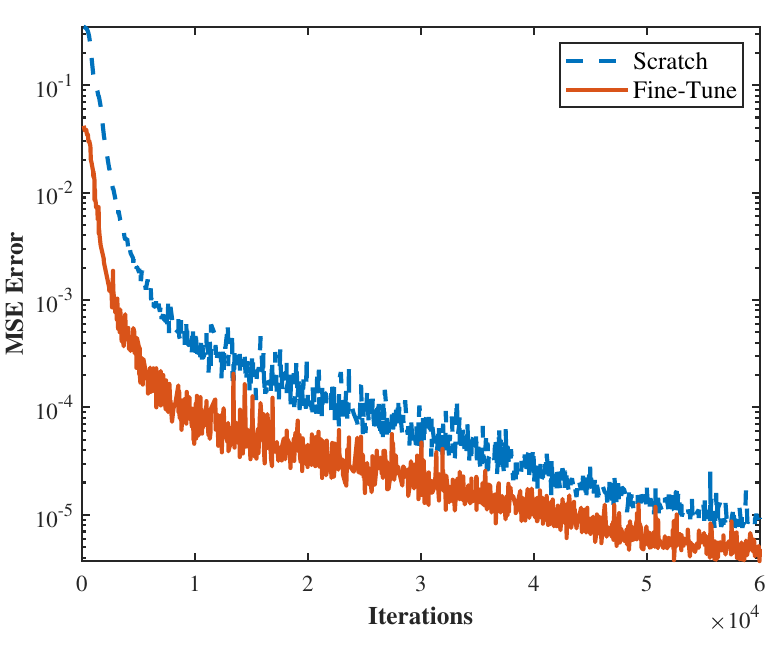}}
    \caption{\label{fig:finetune1}Building models for flow reconstruction: training losses for models trained from scratch via plain supervised learning using $2000$ snapshots and fine-tuned models using $256$ snapshots based on the pre-trained model.}
\end{figure*}

Fig.~\ref{fig:finetune1} displays the training losses for six separate fine-tunings of the pre-trained model for the different Reynolds Numbers. For comparison, we also present the corresponding training losses for models trained from scratch via plain supervised learning using the corresponding first $2000$ snapshots as labeled data. The principal distinction lies in the fact that the parameters of the latter models are initialized randomly. Therefore, at the initial stage the fine-tuned models exhibit much smaller errors than the models trained from scratch. Consequently, the overall training losses by the SSL technique are smaller than that by plain supervised learning. These results come as no surprise for $Re=200$, $300$, $400$ and $500$, 
because these data were already used by the model during the pre-training.  However, for $Re=250$ and $600$, the SSL strategy uses only $256$ pertinent snapshots, while the supervised learning employs $2000$ ones. This demonstrates that the SSL technique is effective to train models targeted at interpolated and extrapolated scenarios of adjacent Reynolds numbers.

\begin{table}[!ht]
    \centering
    
    \begin{tabular}{|c|c|c|c|}
    \hline
        Re & num of input points & MSE (scratch) & MSE (fine tune) \\ \hline
        200 & 354-708 & 3.81E-06 & 1.70E-06 \\ \hline
        250 & 354-708 & 4.66E-06 & 1.90E-06 \\ \hline
        300 & 354-708 & 5.65E-06 & 2.05E-06 \\ \hline
        400 & 354-708 & 5.81E-06 & 2.58E-06 \\ \hline
        500 & 354-708 & 6.47E-06 & 3.12E-06 \\ \hline
        600 & 354-708 & 8.09E-06 & 3.34E-06 \\ \hline
    \end{tabular}
    \caption{Test mean squared errors~(MSEs) for flow reconstructions between models trained from scratch via plain supervised learning and models trained via the SSL. }
    \label{table:mse_reconstruction}
\end{table}
Once the fine-tunings are done, the models are ready to perform the task of flow reconstructions. Firsly, a random number $n$ is drawn from $[354, 708]$. Secondly, $n$ data points are randomly sampled within the observation window as illustrated in Fig.~\ref{fig:sketch_re}, from one snapshot of the test set. Tthe model then reconstructs velocity and pressure fields of the entire snapshot. This precedure is repeated for every $400$ snapshots of each $Re$ in the test set.
TABLE ~\ref{table:mse_reconstruction} presents a comparison of the test MSEs between  the models trained from scratch via plain supervised learning and the models trained via the SSL strategy. Over the six Reynolds numbers considered, the SSL strategy exhibits lower errors than its counterpart for all the test data. Remarkably, even at $Re=250$ and $600$, the models trained via SSL performs much better than the models trained via supervised learning. By leveraging a larger set of data at adjacent Reynolds numbers during the pre-training, the SSL strategy demonstrates superior performance for flow reconstruction, despite the fact that there is no sufficient data for the two specific Reynolds numbers.

\begin{figure*}
    \centering
    \subfloat[Error maps of $p(pa)$, $u(m/s)$ ans $v(m/s)$ at Reynolds number $250$]{\includegraphics[width=1.8\columnwidth]{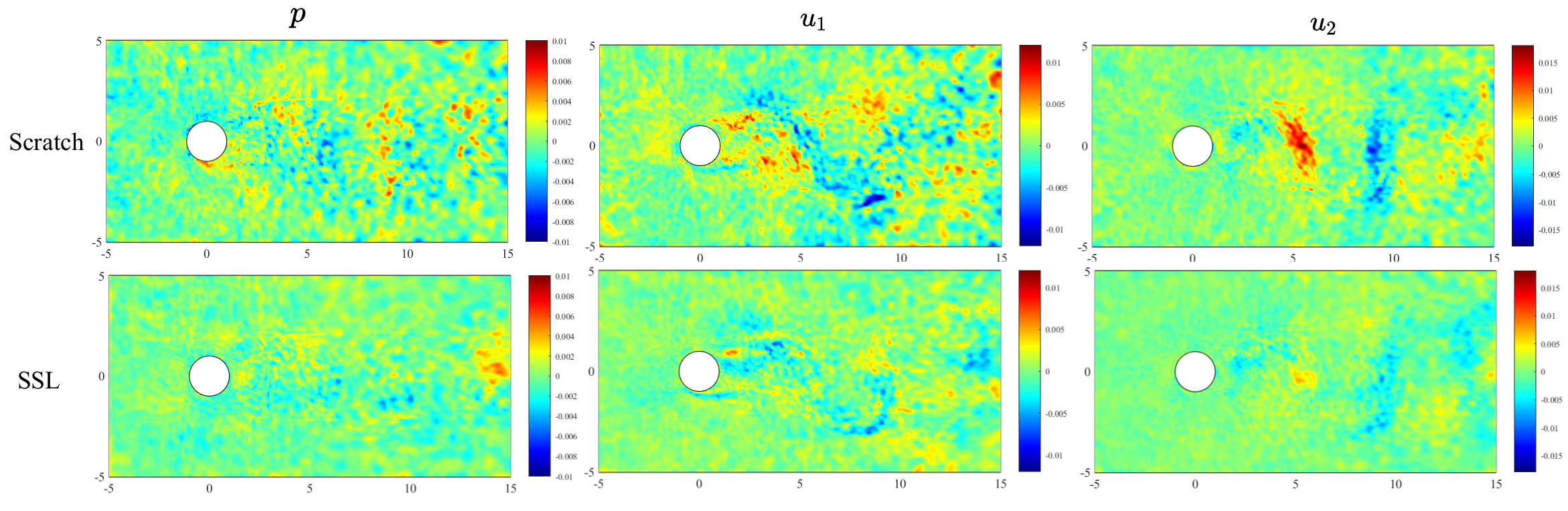}}\hspace{1pt} \\
    
    \subfloat[Error maps of $p(pa)$, $u(m/s)$ ans $v(m/s)$ at Reynolds number $600$]{\includegraphics[width=1.8\columnwidth]{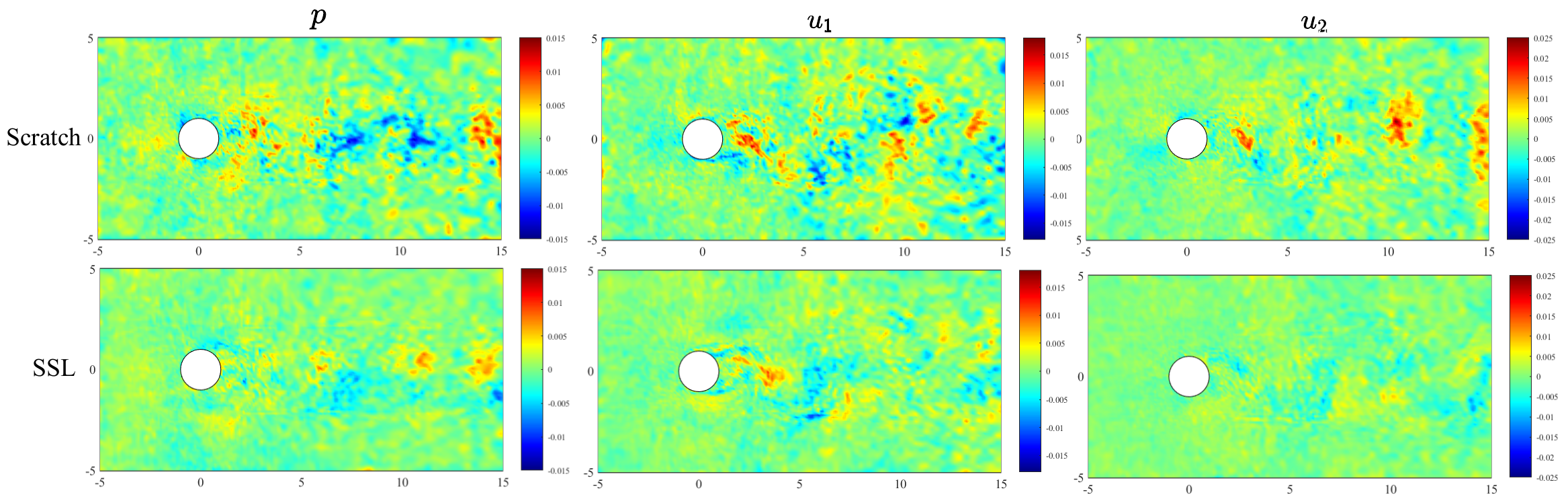}}
    \caption{\label{fig:reErrormap}Error maps of flow reconstruction from models trained from scratch via plain supervised learning and models trained via the self-supervised learning (SSL) at $Re=250$ and $600$.}
\end{figure*}

The error maps for flow reconstruction at $Re=250$ and $600$ from the two types of models are further depicted in Fig.~\ref{fig:reErrormap}, where $p$, $u_1$, and $u_2$ represent the pressure and the two components of the velocity of an arbitrary snapshot in the test set. The error is calculated as $\hat{\mathbf{y}} - \mathbf{y}$, where $\hat{\mathbf{y}}$ denotes the results generated by the models, and $\mathbf{y}$ represents the true values from CFD. The results illustrate that the models trained via the SSL strategy exhibit significantly smaller errors compared to those trained from scratch via plain supervised learning. Specifically, the plain models produce numerous high-frequency noises in the downstream of the cylinder, whreas the models of SSL effectively mitigates these errors, resulting in a substantial enhancement of accuracy for flow reconstruction.

\subsection{Downstream Task 2: Predict progression of flow field}

The product design process in engineering typically relies on resource-intensive and time-consuming high fidelity computer simulations. Various surrogate models have been developed to speed up the process. In this context, we consider the development of surrogate models for flow prediction as another downstream task of the SSL. 

To build the surrogate models, the pre-trained model is separately fine-tuned using supervised learning with labeled data. For each $Re$, $N_2$ consecutive snapshot pairs are randomly selected from the first $2000$ snapshots, which contains $1999$ snapshot pairs, as labeled data. Note that for $Re=200$, $300$, $400$ and $500$, all snapshots have already been used for the pre-training, while the snapshots for $Re=250$ and $600$ have not. Specifically, all~($14,914$) data points of each of the first snapshots in the pairs with three dimensions $(p, u1, u2)$ are taken by the encoder with coordinate information $(x_{1}, x_{2})$ embedded by positional embedding layers. Meanwhile, the same number of coordinates $(x1,x2)$ are provided to the decoder so that a complete snapshot $\Delta t$ ahead in the future is predicted by the Transformer model. The training error is computed as the difference between the predicted values and the true values of the second snapshots in the pairs.

Analogous to the pretext tasks, the learning rate in these fine-tunings follows the one-cycle learning rate policy.  The policy is determined by the initial, maximum and final learning rates set at $3\times 10^{-7}$, $3\times 10^{-4}$, and $3\times 10^{-7}$, respectively. The total number of iterations spans $30,000$. All other training configurations remain consistent with those employed in the pretext task.

\begin{figure*}[t]
    \centering
    \subfloat[MSE loss at Reynolds number 200]{\includegraphics[width=.65\columnwidth]{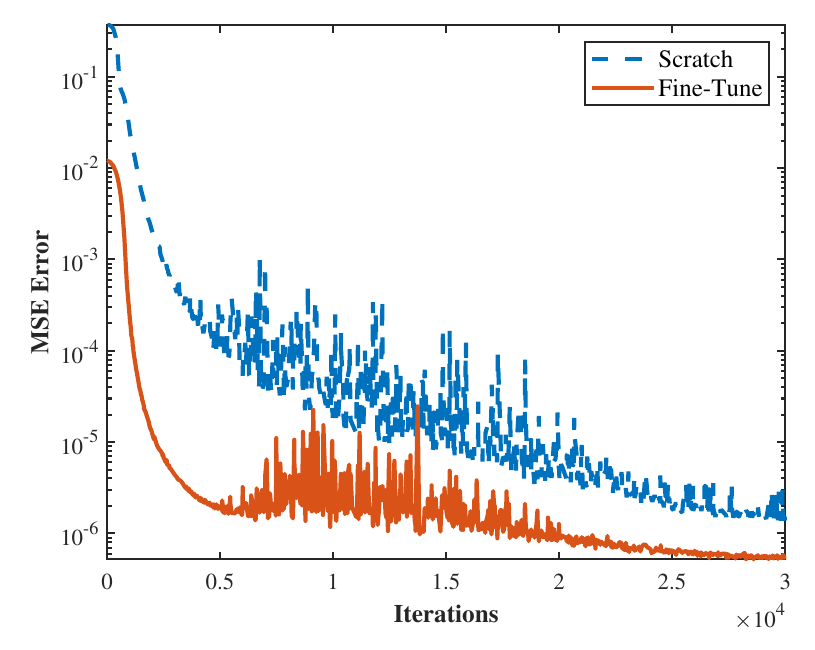}}\hspace{1pt}
    \subfloat[MSE eloss at Reynolds number 250]{\includegraphics[width=.65\columnwidth]{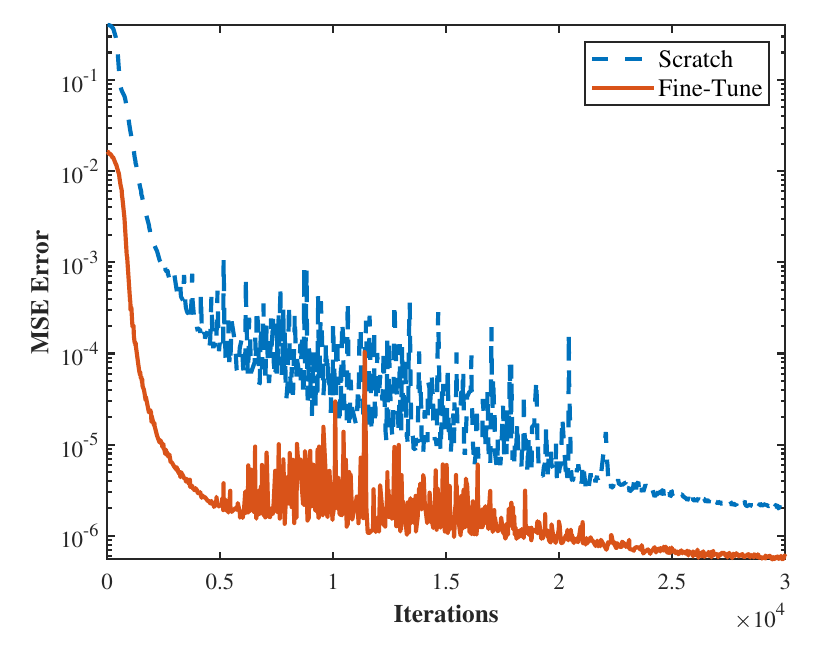}}\hspace{1pt}
    \subfloat[MSE loss at Reynolds number 300]{\includegraphics[width=.65\columnwidth]{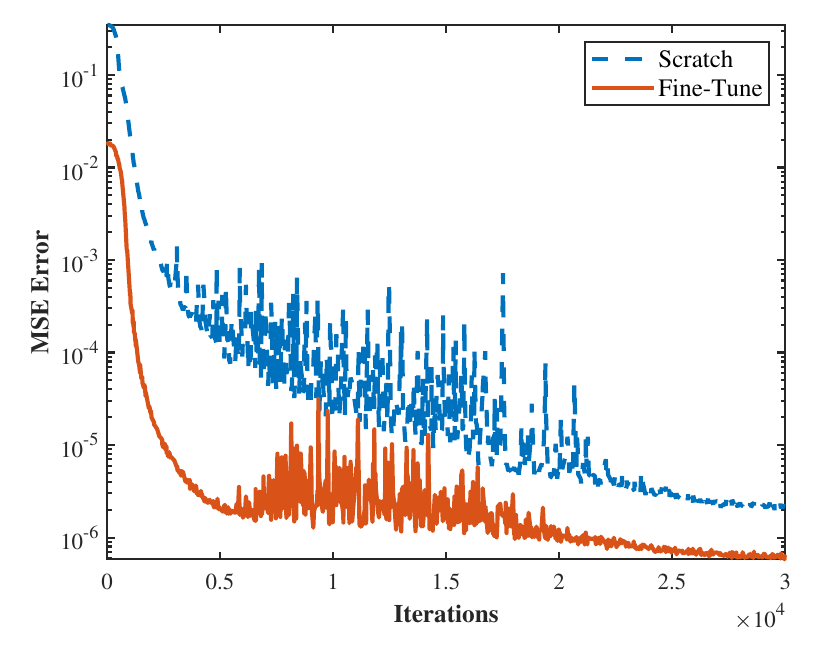}}\\
    \subfloat[MSE loss at Reynolds number 400]{\includegraphics[width=.65\columnwidth]{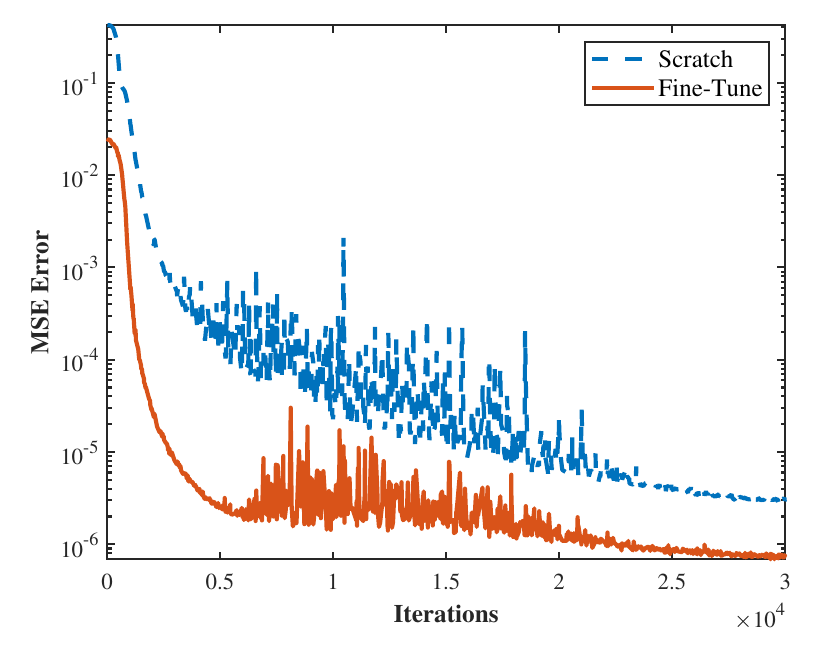}}\hspace{1pt}
    \subfloat[MSE loss at Reynolds number 500]{\includegraphics[width=.65\columnwidth]{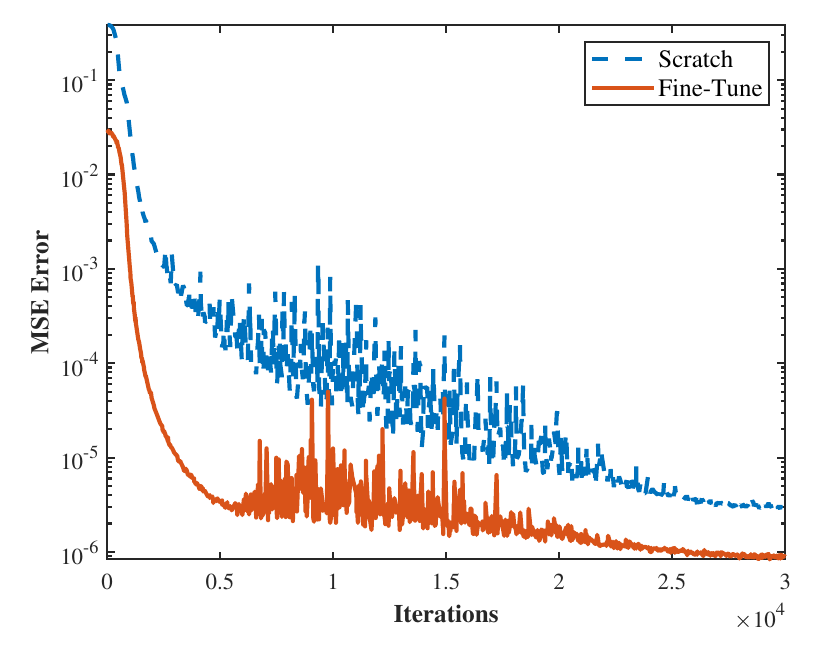}}\hspace{1pt}
    \subfloat[MSE loss at Reynolds number 600]{\includegraphics[width=.65\columnwidth]{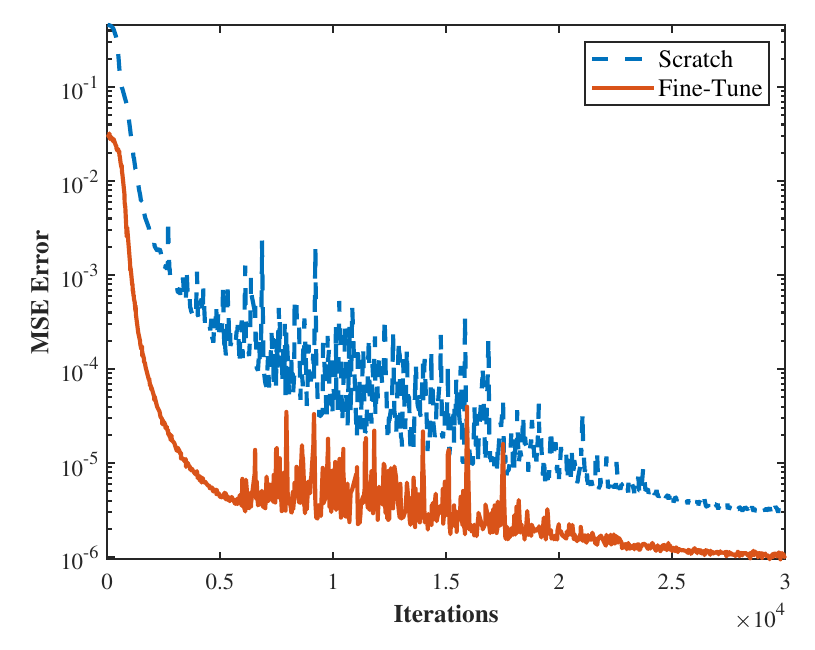}}
    \caption{\label{fig:finetue2}Buiding models for flow prediction: training losses for models trained from scratch via plain supervised learning using randomly chosen $128$ consecutive snapshot pairs and fine-tuned models using randomly chosen $128$ consecutive snapshot pairs based on the pre-trained model.}
\end{figure*}

Fig.~\ref{fig:finetue2} displays the training losses for six separate fine-tunings of the pre-trained model for the different Reynolds numbers. Each fine-tuning is performed via supervised learing using $128$ snapshot pairs randomly selected at the corresponding Reynolds number. Meanwhile, we also show the corresponding training losses for models trained from scratch via plain supervised learning using $128$ randomly selected snapshot pairs. Due to the initialization with random parameters for the networks, the models trained from scratch have larger errors than that of the fine-tuned models. Moreover, the fine-tuned models appear to have a sharp decrease in losses during the first iterations, and therefore the overall training losses using the SSL technique are significantly smaller than those using plain supervised learning. This is even true for $Re=250$ and $600$, data of which are not seen in the pre-trained phase. This again demonstrates that the SSL technique is effective for training models that are aimed at interpolated and extrapolated scenarios of adjacent Reynolds numbers.

\begin{table}[!ht]
    \centering
    \begin{tabular}{|c|c|c|c|}
    \hline
        Re & num of snapshot pairs & MSE (scratch) & MSE (fine tune) \\ \hline
        200 & 128 & 2.40E-06 & 1.46E-06 \\ \hline
        200 & 256 & 2.86E-06 & 1.48E-06 \\ \hline
        200 & 512 & 2.60E-06 & 1.59E-06 \\ \hline
        200 & 1024 & 2.76E-06 & 1.25E-06 \\ \hline
        250 & 128 & 3.00E-06 & 1.55E-06 \\ \hline
        250 & 256 & 3.40E-06 & 1.54E-06 \\ \hline
        250 & 512 & 2.54E-06 & 1.53E-06 \\ \hline
        250 & 1024 & 2.79E-06 & 1.53E-06 \\ \hline
        300 & 128 & 3.39E-06 & 1.68E-06 \\ \hline
        300 & 256 & 3.61E-06 & 1.66E-06 \\ \hline
        300 & 512 & 3.63E-06 & 1.64E-06 \\ \hline
        300 & 1024 & 3.52E-06 & 1.64E-06 \\ \hline
        400 & 128 & 3.89E-06 & 1.97E-06 \\ \hline
        400 & 256 & 3.81E-06 & 1.95E-06 \\ \hline
        400 & 512 & 3.87E-06 & 1.94E-06 \\ \hline
        400 & 1024 & 4.11E-06 & 1.94E-06 \\ \hline
        500 & 128 & 4.94E-06 & 2.29E-06 \\ \hline
        500 & 256 & 4.78E-06 & 2.30E-06 \\ \hline
        500 & 512 & 4.69E-06 & 2.28E-06 \\ \hline
        500 & 1024 & 5.09E-06 & 2.27E-06 \\ \hline
        600 & 128 & 4.43E-06 & 2.69E-06 \\ \hline
        600 & 256 & 5.46E-06 & 2.65E-06 \\ \hline
        600 & 512 & 4.49E-06 & 2.65E-06 \\ \hline
        600 & 1024 & 4.41E-06 & 2.63E-06 \\ \hline
    \end{tabular}
    \caption{Test mean squared errors~(MSEs) for flow prediction between models trained from scratch via plain supervised learning and models trained via the SSL. The second column indicates the numbers of consecutive snapshot pairs used in the plain supervised learning and also in the fine-tuning stage of the SSL.}
    \label{table:mse_prediction}
\end{table}

Once fine-tunings are finished, the models are ready to perform the task of flow predictions. TABLE ~\ref{table:mse_prediction} presents an overview of the test errors associated with models trained from scratch via plain supervised learning and the models trained via the SSL. To make a comprehensive comparision, we vary the number of snapshot pairs (among $128$, $256$, $512$ and $1024$) used in the plain supervised learning and during the fine-tuning stage of the SSL. There is no monotonical decrease in the test errors for both types of models when the number of snapshot pairs is increased. However, the test errors from the models trained via the SSL generally are
much smaller than those of the models trained via plain supervised learning, irrepsective of different Reynolds numbers or different numbers of snapshot pairs used for training. These results are remarkable, as {\it no temporal information was considered during the pre-training stage of the SSL}. When only $128$ pairs of causual relations are taken into account in the fine-tuning stage, the models trained via the SSL already significantly outperform their counterparts informed by $1024$ pairs. This is true even for interpolated and extrapoloated Reynolds numbers of $Re=250$ and $600$, data of which were not seen at all during the pre-training stage of the SSL.

\begin{figure*}
    \centering
    \subfloat[Error maps of surrogate models at Reynolds number 250]{\includegraphics[width=1.8\columnwidth]{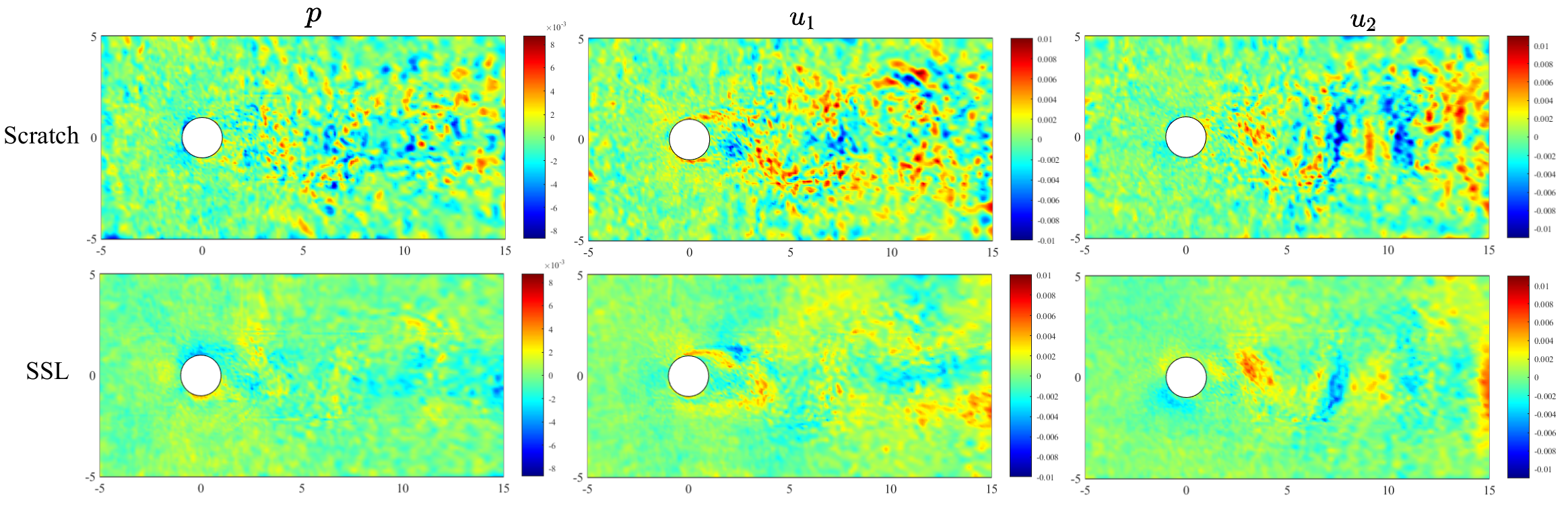}}\hspace{1pt} \\
    
    \subfloat[Error maps of surrogate models at Reynolds number 600]{\includegraphics[width=1.8\columnwidth]{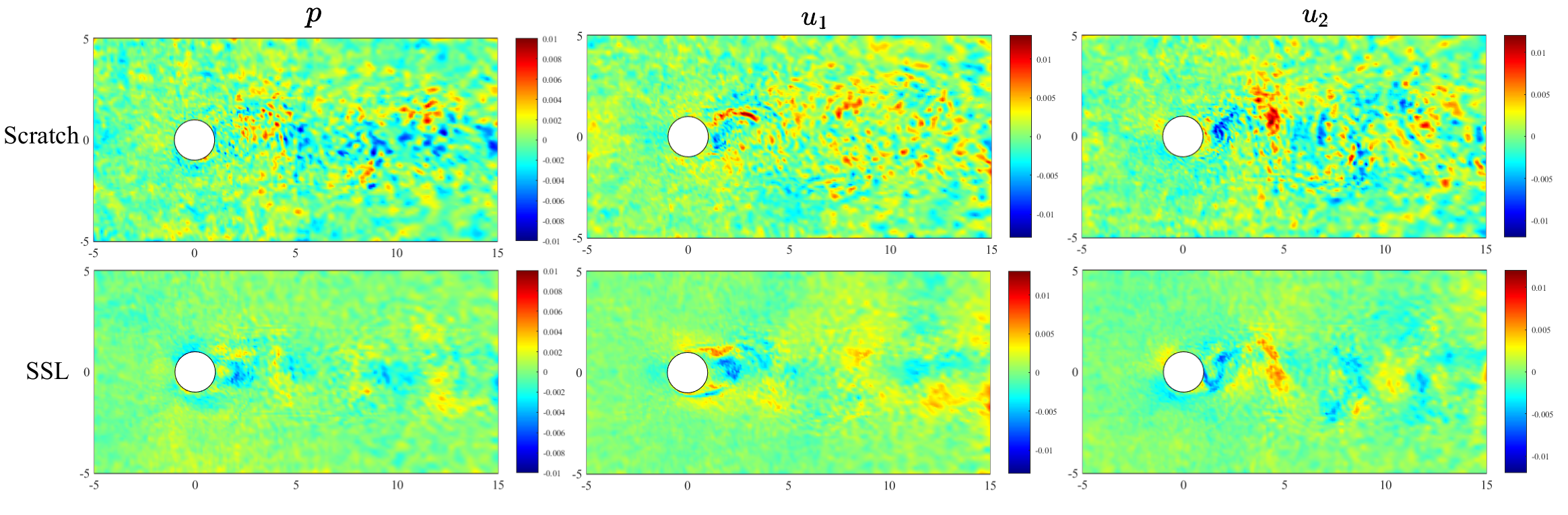}}
    \caption{\label{fig:preErrormap}Error maps of flow prediction from models trained from scratch via plain supervised learning and models trained via the self-supervised learning~(SSL) at $Re=250$ and $600$. These correspond to the two rows with $128$ snapshot pairs in TABLE~\ref{table:mse_prediction}.}
\end{figure*}

The error maps of flow prediction at $Re=250$ and $600$ from the two types of models are shown in Fig.~\ref{fig:preErrormap}. These correspond to the two rows with $128$ snapshot pairs in TABLE~\ref{table:mse_prediction}. We observe that the errors from the models trained from scrtach concentrate within the the downstream wake of the cylinder, where high-frequency noises are present. The results from the models trained via the SSL demonstrate superior performance with relatively smaller errors.

\begin{figure}[h]
\includegraphics[width=0.9\columnwidth]{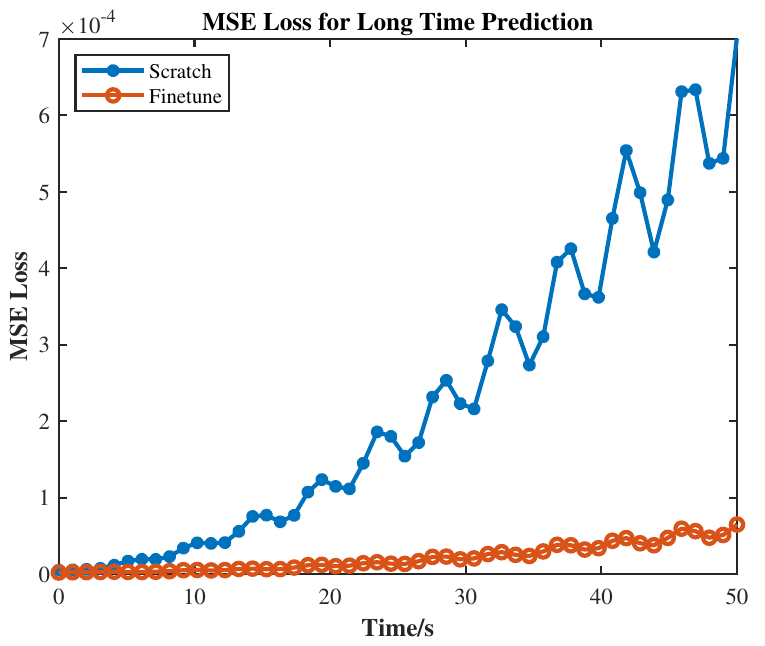}
\caption{\label{fig:fig11} The errors of long-term prediction task from the fine-tuned model via the SSL and the surrogate model trained from scratch via plain supervised learning: $Re=500$.}
\end{figure}

In addition, the capacity for accurate long-term predictions stands as a pivotal evaluation criterion for surrogate models. We examplify the long-term prediction errors at $Re=500$ in Fig.~\ref{fig:fig11}, where an output snapshot of the surrogate model is employed as input again for successive predictions.  While the difference between the two kinds of models are not substantial for the case of a single-step prediction, the error indeed accumulates fast and diverge apparently in long time. In particular, after $50$ prediction cycles, a striking disparity becomes evident: the model trained from scratch experiences rapid error accumulation, with errors becoming six times larger than those observed with the SSL-trained model. This result is particularly intriguing, given that the pre-training process within the SSL relies solely on snapshots without using temporally causal information.

\section{Conclusion and perspective}
\label{sec_conclusion}

Although the self-supervised learning~(SSL) strategy has gained prominence in the training of large models, its application in the field of fluid mechanics has been unexplored. In this study, we conduct a preliminary investigation of the SSL technique for flow reconstruction and prediction. Specifically, we design a pretext task in which a portion of each snapshot in a large dataset is randomly masked. The Transformer model is then pre-trained to recover the complete snapshots without discriminating the data between Reynolds numbers. Subsequently, the pre-trained mode is separately fine-tuned with a small amount of data from specific Reynolds numbers for two types of downstream tasks, that is, flow reconstruction and prediction. The first fine-tuned models are capable of accurately reconstructing the entire flow field from a small amount of data available within a confined observation window in space and time. This is true even for $Re=250$ and $600$, for which no data were available during the pre-training phase. The second fine-tuned models are able to correctly predict the evolution of flow fields over many cycles of time periods. They significantly reduce the cumulative errors, even though the timestamps of the data were not used as labels during the pre-training phase. Compared to models trained from scratch using simple supervised learning, the SSL strategy shows remarkable improvements for both types of tasks. These results demonstrate that the SSL technique can leverage a large amount of unlabeled data to improve the model generalization and therefore, provides transfer learning capability to a certain degree.

Our research suggests other potential applications of the SSL technique in the field of fluid mechanics. For example, after pre-training with a large amount of data from small Reynolds numbers and fine-tuning with a small amount of data at one large Reynolds number, the model may be able to accurately reconstruct the flow field of the later. For transient flows, different datasets exist as a sequence of snapshots, but with different time steps. These datasets cannot be directly combined to train a surrogate model with a fixed time step. However, they can be effectively employed together in the pretext task during the pre-training phase using the SSL technique.
Finally, a model can be pre-trained with flow fields covering different shapes of a blunt body, and then fine-tuned to predict flow over a new blunt body. This could potentially speed up the design process of structure-shape optimization in flows.

\begin{acknowledgments}
B. Xu is partially supported by the Post-doctoral Fellowship of Zhejiang University.
X. Bian received the starting grant from 100 Talents Program of Zhejiang University.
\end{acknowledgments}

\section*{Data Availability Statement}
All of the codes for the neural network models and data generation are available in the supplementary materials.

\section{Appendixes}
The relative position and absolute position are usually both important for physical systems. The Rotary Positional Embedding (RoPE) proposed by Su et al~\cite{Rope} is a new type of position encoding that unifies absolute and relative approaches. By taking query $q$ as an example, the embedding function $f(\mathbf{q}, m)$ can be written as:
\begin{equation*}
    f(\mathbf{q}, m)=\left(\begin{array}{cccc}
M_{1} & & & \\
& M_{2} & & \\
& & \ddots & \\
& & & M_{d / 2}
\end{array}\right)\left(\begin{array}{c}
q_{1} \\
q_{2} \\
\vdots \\
q_{d}
\end{array}\right)=\boldsymbol{\Theta}_{\mathbf{m}} \mathbf{Q}_{\mathbf{m}}=\boldsymbol{\Theta}_{\mathbf{m}} \mathbf{W}_{\mathbf{q}} \mathbf{X}_{\mathbf{m}}
\end{equation*}

where, $\begin{pmatrix}  \cos m \theta_j  & -\sin m \theta_{j} \\ \sin m \theta_{j}  & \cos m \theta_{j} \end{pmatrix}$, $\boldsymbol{\Theta}_{\mathbf{m}}$ is the block diagonal matrix, $\mathbf{W}_{q}$ is the learnable query weight, and $\mathbf{X}_{m}$ is the embedding of the token with position $m$. $\theta_{j}$ is set to $1000^{-2(j-1)/d}$, and $d$ is the dimension of Transformer model. Again, we also have the corresponding equation for key array $\mathbf{k}$ and value array $\mathbf{v}$. 

With relative ease RoPE can be extended into the multidimensional case~\cite{rope-eleutherai}. To represent two dimensions, two independent 1-dimensional rotary embeddings can be used. To implement this, we can split each of $\mathbf{q}$ and $\mathbf{k}$ in half and apply rotary piece-wise as follows:

\begin{align*}
& \langle f(\mathbf{q}, m, i),f(\mathbf{k}, n, j) \rangle \\
&= \langle f_1(\mathbf{q}_{:d/2}, m),f_1(\mathbf{k}_{:d/2}, n) \rangle + \langle f_2(\mathbf{q}_{d/2:}, i),f_2(\mathbf{k}_{d/2:}, j) \rangle\\
&= g_1(\mathbf{q}_{:d/2}, \mathbf{k}_{:d/2}, m - n) + g_2(\mathbf{q}_{d/2:}, \mathbf{k}_{d/2:}, i - j)\\
&= g(\mathbf{q}, \mathbf{k}, m - n, i - j) 
\end{align*}

This formulation can also be further extended to data of an arbitrary number of dimensions.

\nocite{*}
\bibliography{main}

\end{document}